\documentclass{nature}

\title{Spectroscopy of the superconducting proximity effect in nanowires using integrated quantum dots}

\author{Christian J\"unger$^{1}$, Andreas Baumgartner$^{1}$, Rapha\"elle Delagrange$^1$,  Denis Chevallier$^1$, Sebastian Lehmann$^2$, Malin Nilsson$^2$, Kimberly A. Dick$^{2,3}$, Claes Thelander$^2$ \& Christian Sch\"onenberger$^1$}

\usepackage{graphicx}
\makeatletter
\let\saved@includegraphics\includegraphics
\AtBeginDocument{\let\includegraphics\saved@includegraphics}
\renewenvironment*{figure}{\@float{figure}}{\end@float}
\makeatother

\begin{document}

\maketitle

\begin{affiliations}
 \item Department of Physics, University of Basel, Klingelbergstrasse 82, CH-4056 Basel, Switzerland
 \item Division of Solid State Physics and NanoLund, Lund University, S-221 00 Lund, Sweden
 \item Center for Analysis and Synthesis Lund University, S-221 00 Lund, Sweden
\end{affiliations}

\begin{abstract}
The superconducting proximity effect has been the focus of significant research efforts over many years and has recently attracted renewed interest as the basis of topologically non-trivial states in materials with a large spin orbit interaction, with protected boundary states useful for quantum information technologies. However, spectroscopy of these states is challenging because of the limited spatial and energetic control of conventional tunnel barriers. Here, we report electronic spectroscopy measurements of the proximity gap in a semiconducting indium arsenide (InAs) nanowire (NW) segment coupled to a superconductor (SC), using a spatially separated quantum dot (QD) formed deterministically during the crystal growth. We extract the characteristic parameters describing the proximity gap which is suppressed for lower electron densities and fully developed for larger ones. This gate-tunable transition of the proximity effect can be understood as a transition from the long to the short junction regime of subgap bound states in the NW segment. Our device architecture opens up the way to systematic, unambiguous spectroscopy studies of subgap bound states, such as Majorana bound states.\\
\end{abstract}
\section*{\label{sec:level1}Introduction}
Coupling a superconductor (SC) to a metal or a semiconductor results in the so called superconducting proximity effect in the normal material\cite{Beenakker1992}. If the proximitised material is a low-dimensional semiconductor, this phenomenon can, for example, be used as a source of spin-entangled electrons\cite{Hofstetter2009,Schindele2012} or superconducting magnetometers\cite{Cleuziou2006}. By combining an s-wave SC with a one-dimensional semiconducting nanowire (NW) with large spin orbit interaction, one can artificially create a proximity region with superconductivity of p-wave character. This can give rise to exotic quantum states at the ends of the SC, such as Majorana bound states\cite{Mourik2012,Albrecht2016,Deng2016,Zhang2018}, potentially useful as building blocks for topological quantum information processing\cite{Alicea2011,Nayak2008}. However, an unambiguous characterization of superconducting subgap states and the proximity region in NWs remains challenging. Several theoretical proposals suggest to use a quantum dot (QD) as a spectrometer to investigate Majorana bound state lifetime\cite{Leijnse2011,Hoffman2017}, spin texture\cite{Chevallier2018} or parity\cite{Gharavi2016}. A first experiment was reported recently in which the QD was defined by electrical gating\cite{Deng2016}. However, a systematic and deterministic spectroscopy, requires a spatially well defined, weakly coupled, QD with sharp tunnel barriers across the complete NW, which does not hybridise with the bound states or the SC under investigation.\\
Here, we introduce a new material platform that allows us to perform ideal tunnel spectroscopy: we use an in-situ grown axial QD in an InAs NW covering the complete diameter of the NW, to probe the superconducting proximity region of a NW segment close to the superconducting contact. The QD is defined by two potential barriers that form when the NW crystal structure is changed from zincblende (ZB) to wurtzite (WZ), which can be achieved with atomic precision by controlling the growth parameters\cite{Dick2010,Lehmann2013}. These QDs are electrically and spatially well defined, which allows us to probe the induced gap at a precise distance from the QD and with predictable coupling parameters. In-situ grown barriers have previously been used to investigate double QD physics\cite{Rossella2014,Nilsson2018}. Here, we use such artificial QDs to study the evolution of the proximity induced superconductivity in a controlled and systematic manner. We demonstrate this type of spectroscopy using two different transport regimes of the QD: in the cotunneling regime, where the QD can be seen as a single tunnel barrier, and in the sequential tunneling regime, where the QD acts as an energy filter with Coulomb blockade (CB) resonances. The complementary measurements allow us to draw a clear picture of how the proximity gap forms in a NW segment, for which we present an intuitive explanation supported by numerical simulations.
\begin{figure}
	\begin{center}
	\includegraphics[width=0.8\columnwidth]{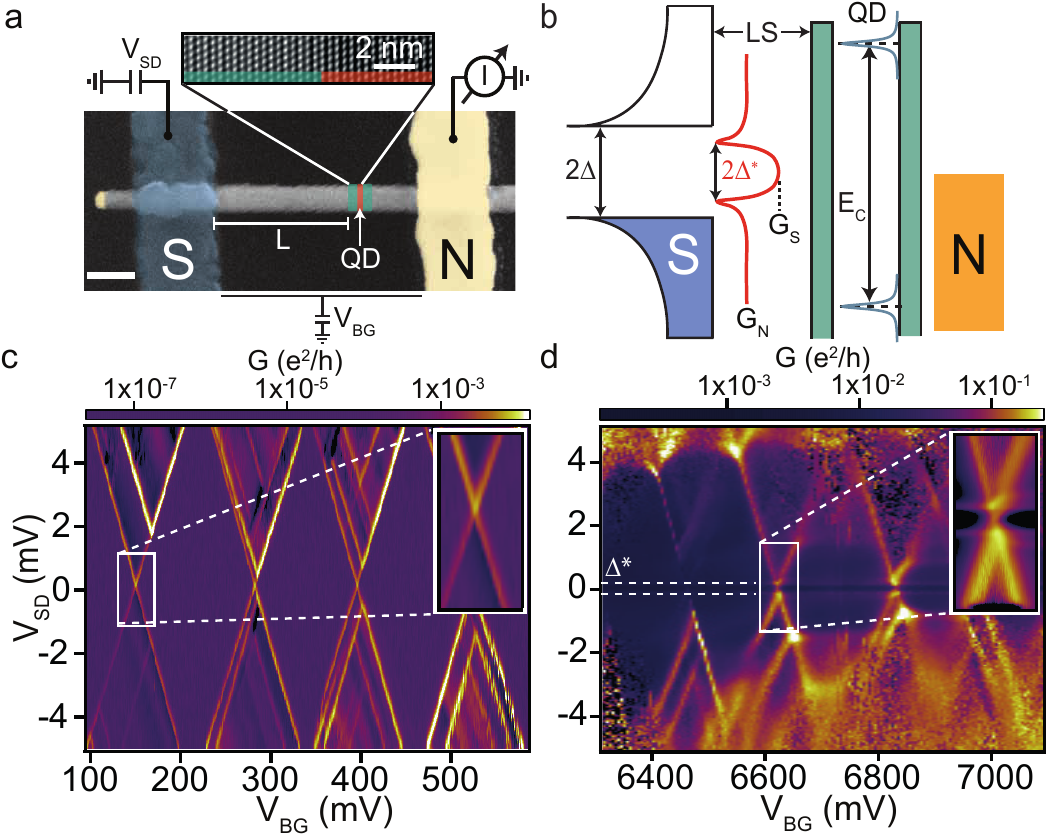}
    \end{center} 	
	\caption{Crystal phase engineered QD in an S - N - QD device. \textbf{a} False color scanning electron micrograph of the investigated device, consisting of a superconductor S (blue) - InAs NW - normal metal N (yellow) junction (scale bar: \SI{100}{nm}). A QD (red) forms between two in-situ grown tunnel barriers (green) in the WZ phase. The measurement scheme is shown schematically. The inset shows a high resolution transmission electron micrograph of the atomically sharp ZB/WZ interface of a comparable NW. \textbf{b} Energy diagram of such a system with an illustration of the proximity gap $\Delta\textsuperscript{*}$ in the lead segment (LS). \textbf{c} Differential conductance $G$ as a function of the backgate voltage $V_{BG}$ and the source-drain bias $V_{SD}$ in the superconducting state. The inset shows a single CB resonance in more detail. \textbf{d} CB diamonds at higher backgate voltages exhibiting a gap in the transport characteristics.}
	\label{fig:setup}
\end{figure}
\section*{\label{sec:level2}Results}
\subsection{\label{subsec:level1}Device and characterization.}
We use an InAs NW with an in situ grown axial QD formed by controlling its crystal phase structure in the growth direction\cite{Lehmann2013}: two thin segments (\SI{30}{\nano \meter}) of WZ phase are grown in the otherwise ZB NW. These segments act as atomically precise hard-wall tunnel barriers for electrons, because the ZB and WZ bandstructure align with a conduction band offset of \SI{\sim100}{\milli eV}\cite{Nilsson2016,Chen2017}. Consequently, in the ZB section (\SI{20}{\nano \meter}) between the two barriers a QD is formed\cite{Nilsson2016}. A false color scanning electron micrograph of the actual device is shown in Fig.\ref{fig:setup}a, together with a transmission electron micrograph of the atomically precise interface in a representative NW in the inset. The superconducting electrode consists of titanium/aluminum (Ti/Al) and the normal contact of titanium/gold (Ti/Au), fabricated by electron beam lithography. The differential conductance $ G = dI/dV_{SD}$ as a function of the backgate voltage $V_{BG}$ in the normal state (in an external magnetic field of $B$ = \SI{50}{\milli \tesla}) shows regularly spaced CB resonances (see supplementary material Fig.\ref{fig:GammaGate}a) for which  we find a systematic increase of the total  tunnel coupling $\Gamma$ with increasing $V_{BG}$ (see Fig.\ref{fig:GammaGate}b), consistent with the lowering of the tunnel barrier when the bandstructure is shifted to lower energies with respect to the Fermi energy (see schematic in Fig.\ref{fig:EnergySchemeFermi})\cite{Nilsson2016}.\\
In the device discussed here, the QD is located about L\SI{\approx 350}{\nano \meter} from the Al electrode, leaving a bare NW segment of this length between the QD and the SC, which we refer to as the ``lead segment'' (LS). This provides a new experimental situation, as the QD is not directly coupled to the SC (as it might be the case in other experiments\cite{Doh2008,Gramich2015,Gramich2017}) and can be used to probe the LS. $V_{BG}$ here directly tunes the chemical potential in both the LS and the QD.
We record $G$ as a function of the bias voltage $V_{SD}$ applied between the SC and the normal metal contact. If the electrons would tunnel directly from the QD to the superconducting electrode, we would expect to see a gap similar to  the one of the bulk $\Delta$\SI{\approx 210}{\micro eV} \cite{Court2007} of the SC, independent of $V_{BG}$. Since most of the bias drops over the QD, $G$ is proportional to the density of states (DOS) in the LS. This is the case for positive $V_{BG}$ where carriers accumulate in the NW. Therefore we can perform spectroscopy on the LS by tunneling from the QD (see Fig.\ref{fig:setup}b).\\
An interesting transition in the conductance can already be found in the overview data in Fig.\ref{fig:setup}c,d presenting regular CB diamonds in the superconducting state (charging energy $E_{c} \approx$ \SI{6}{\milli eV}, level spacing $\epsilon \approx$ \SIrange{1.5}{2}{\milli eV}) for two different regimes of $V_{BG}$. While for positive gate voltages $V_{BG}\gtrsim$ \SI{3 }{V} we observe a superconducting gap around zero bias (c.f. Fig.\ref{fig:setup}d), for low gate voltages (c.f. Fig.\ref{fig:setup}c) we do not find any features related to superconductivity, but rather the standard sequence of diamonds.
\subsection{\label{sec:level3}Measurements.}
\begin{figure}
	\begin{center}
	\includegraphics[width=0.75\textwidth,height=\textheight,keepaspectratio]{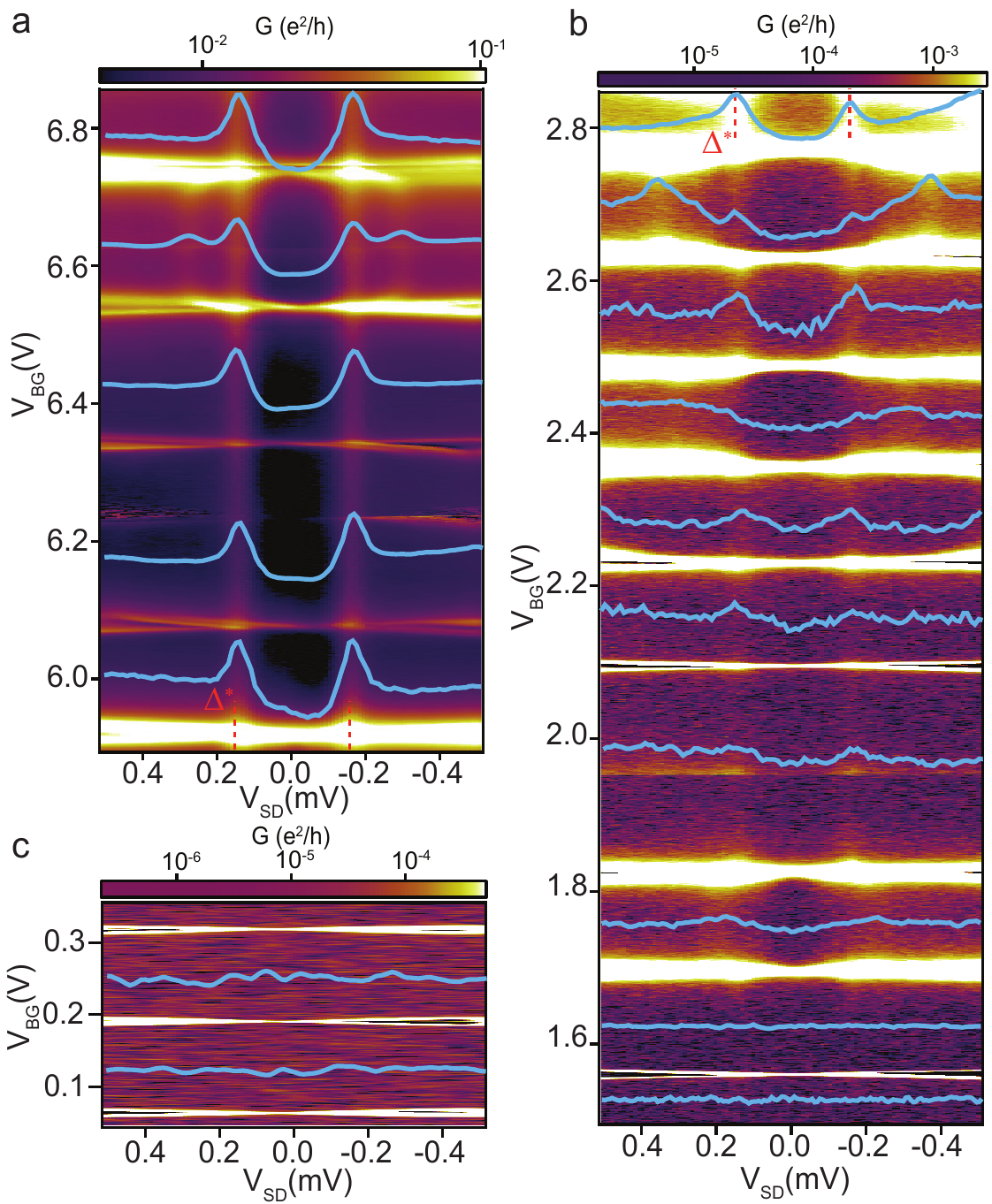}
    \end{center} 	
	\caption{\label{fig:Cotunnel}Proximity gap probed by cotunneling spectroscopy. Differential conductance $G$ as a function of $V_{BG}$ and $V_{SD}$ for \textbf{a} high, \textbf{b} middle and \textbf{c} low backgate voltages $V_{BG}$. Blue lines are averaged over 20 cross sections taken in the center of the CB diamonds. The cross sections shown in \textbf{c} are averaged over the CB region between the Coulomb resonances.}
\end{figure}

\begin{figure}
	\begin{center}
	\includegraphics[width=0.87\textwidth,height=\textheight,keepaspectratio]{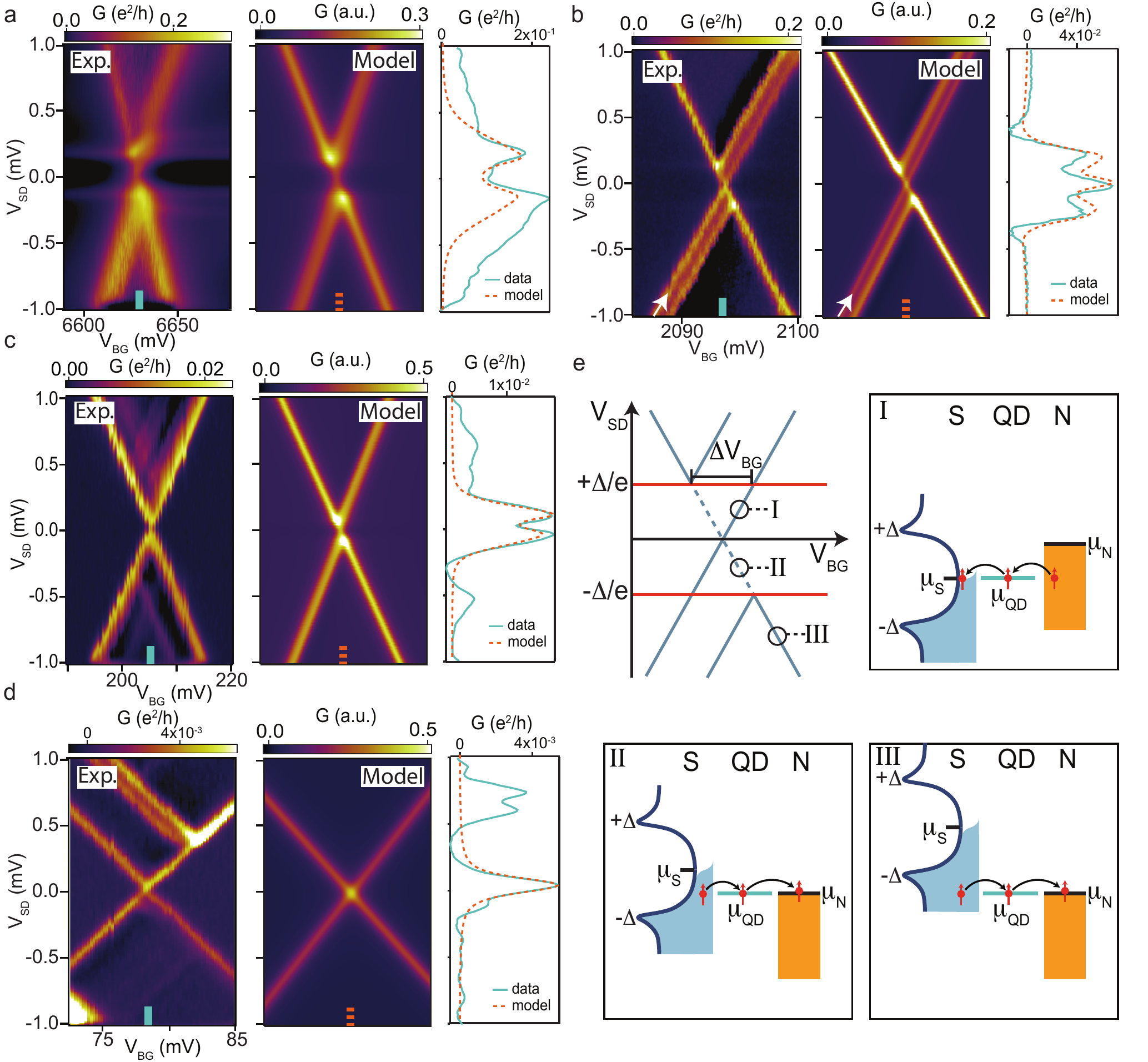}
    \end{center} 	
	\caption{\label{fig:Coulomb}Proximity gap in the resonant tunneling regime. \textbf{a-d} Experiment (left) and resonant tunneling model (middle) and cross sections of both (right) for different CB resonances. Light blue and orange lines indicate the backgate voltage for the cross sections. \textbf{e} Relative positions of the electrochemical potentials of the normal contact ($\mu_{N}$), the SC ($\mu_S = \mu_{N}  -eV_{SD}$) and the QD ($\mu_{QD}$) for selected points in a charge stability diagram. The tips of the diamonds are shifted in gate voltage by $\Delta V_{BG} = 2\Delta\textsuperscript{*}/(\beta e)$ with $\beta$ the lever arm of the CB resonance. In \textbf{a} the tunnel coupling is almost a factor of 3 larger than in \textbf{b-d}. In the model, the tunnel coupling $\Gamma$ is essentially constant for \textbf{b-d}, while the value for the induced superconducting energy gap varies from $\Delta\textsuperscript{*}$=\SI{165}{\micro eV} in \textbf{a}, $\Delta\textsuperscript{*}$=\SI{145}{\micro eV} in \textbf{b},$\Delta\textsuperscript{*}$=\SI{85}{\micro eV} in \textbf{c} and $\Delta\textsuperscript{*} <$\SI{10}{\micro eV} in \textbf{d}.
	}
\end{figure}

The QD can be used as a spectrometer in two different regimes. In the first regime, the QD is kept deep in the CB regime where the charge is fixed for bias voltages in the range of the proximity induced gap $\Delta\textsuperscript{*}$, since $\Delta\textsuperscript{*} << E_c$. In this regime transport is mediated by cotunneling, which is a second order process involving the virtual occupation of a QD state\cite{Franceschi2001}. Here, the QD can be thought of as a single tunnel barrier. In the second regime, the QD electrochemical potential is tuned to to a CB resonance, with transport occurring as first order sequential tunneling. In the following, we discuss the experiments in the two regimes one after the other.\\
First, we investigate the cotunneling regime, where we can understand the SC - LS as an S-N junction which is weakly coupled to the QD. The measured spectrum is shown in Fig.\ref{fig:NormGap}a. It can be characterized by four quantities: the magnitude of the observed gap $\Delta\textsuperscript{*}$, the full width at half the maximum (FWHM) of the peaks at the gap edge, the normal state conductance $G_{N}$ measured at a bias $|V_{SD}|>\Delta\textsuperscript{*}/e$ and $G_S$ the conductance at $V_{SD} = 0$. Since the conductance is not fully suppressed at zero bias, we characterise this softness by defining the suppression factor $S = G_S/G_N $.\\
Figures \ref{fig:Cotunnel}a-c show $G$ as a function of $V_{BG}$ and $V_{SD}$ for three different backgate regimes. Cross sections (Fig.\ref{fig:Cotunnel}c: averaged over 20 cross sections in the CB diamond centers) at different $V_{BG}$ in the CB regime are plotted as blue lines. 
At large gate voltages ($V_{BG}\approx$ \SI{7}{V}, Fig.\ref{fig:Cotunnel}a) we find a clear gap around zero bias, that can be detected down to $V_{BG}\approx$ \SI{2.6}{\volt} (Fig.\ref{fig:Cotunnel}b). The same cross sections normalized to $G_N$ are shown in the supplementary section, Fig.\ref{fig:NormGap}a.
All cross sections exhibit a suppression of $G$ at $V_{SD} = 0$ and $\Delta\textsuperscript{*}\approx$ \SI{150}{\micro eV} (position of the peak maxima). The FWHM and $S$ are shown to be roughly constant over the investigated gate range, see Fig.\ref{fig:NormGap}b, with $FWHM \approx$ \SI{65}{\micro eV}$\pm$\SI{10}{\micro eV} and $S\approx$ $0.5\pm0.1$, respectively. In an ideal SC and weak coupling, $S$ should be close to zero for a strong tunnel barrier. Larger values are often observed for proximity induced gaps, referred to as soft gap\cite{Doh2005,Sand-Jespersen2007,Jellinggaard2016}. For lower gate voltages $V_{BG}<$\SI{2.6}{\volt} (Fig.\ref{fig:Cotunnel}b) it is difficult to perform this analysis because the cotunneling signal is very low, compared to the noise floor of the experiment ($G\approx 10^{-5} e^{2}/h$). However, we can still observe broad peaks down to $V_{BG}\approx$ \SI{1.8}{\volt}. For $V_{BG}<$\SI{1.8}{\volt} the differential conductance is too low to observe any signatures of superconductivity with the resolution of our measurement. For even lower gate voltages $V_{BG}\approx$ \SI{0.2}{\volt} (Fig. \ref{fig:Cotunnel}c) no features inside the CB diamonds can be resolved anymore. In summary, in the regime accessible by the cotunneling experiments, the proximity gap characteristics are roughly independent of the gate voltage.\\
To extract characteristics of the LS for a larger gate range than in the cotunneling regime, we now use the QD resonances (first order process through the QD). The left figure (``Exp.'') of each panel in Fig. \ref{fig:Coulomb}a-d shows detailed measurements of a CB resonance in different backgate regimes, while the central panel shows a simulated map discussed below, and the right panel shows a cross section at $V_{BG}$ as indicated in the left and central panels. In a S-QD-N system, the Coulomb diamond pattern is expected to be affected by superconductivity in the way represented in Fig. \ref{fig:Coulomb}e. The tips of the Coulomb diamonds
are expected to be shifted by $2\Delta\textsuperscript{*}/e$ in bias and by $\Delta V_{BG} = 2\Delta\textsuperscript{*}/\beta e$ (with $\beta$ the lever arm of the QD) in gate voltage \cite{Gramich2015,Gramich2016}.\\
From the QD characteristics in the normal state (see Fig. \ref{fig:GammaGate}a), we find that the tunnel coupling $\Gamma$ increases significantly for $V_{BG}>$ \SI{2.5}{V}. For large $\Gamma$, e.g. at $V_{BG}\approx$ \SI{6.6}{V}, a CB pattern is shown in Fig.\ref{fig:Coulomb}a.
The CB resonances are broad, but clearly show a suppressed conductance around zero bias. We find a gap of $\Delta\textsuperscript{*}\approx$ \SI{150}{\micro eV}, confirming the value obtained in the cotunneling regime (Fig. \ref{fig:Cotunnel}).\\
A CB resonance with a smaller $\Gamma$ value at $V_{BG}\approx$ \SI{2.1}{V} is shown in Fig. \ref{fig:Coulomb}b, which is similar to the one in Fig. \ref{fig:Coulomb}a, but with an additional resolved resonance. As expected (see Fig. \ref{fig:Coulomb}e), the "CB diamond tips" are shifted in energy by $\pm\Delta^*/e$ and $\Delta V_{BG}$ in gate voltage, yielding a consistent value of \linebreak$\Delta^*\sim$ \SI{150}{\micro eV}. We observe an additional resonance that crosses through the gap (white arrows). This line corresponds to the alignment of the Fermi level of the two reservoirs with the QD state. We attribute this to the tunneling through the non-zero DOS remaining at zero bias (see position $\text{I}$ in Fig. \ref{fig:Coulomb}e).\\
At $V_{BG}\approx$ \SI{0.2}{V} the conductance suppression is significantly reduced, as shown in Fig.\ref{fig:Coulomb}c. The CB diamond tips appear to be only slightly separated and shifted. We note that in this gate range we cannot resolve any signal in the cotunneling spectrum, as discussed in Fig. \ref{fig:Cotunnel}.\\
At even lower gate voltages, e.g. $V_{BG}\approx$ \SI{80}{\milli V} (Fig. \ref{fig:Coulomb}d), we do not observe any influence of the superconducting contact, but a regular CB resonance, as found in the normal state.\\
To extract the characteristic numbers from these data, we use a resonant tunneling model for a S - QD - N junction. The current is then given by\cite{Yeyati1997,Gramich2015}
$I = \int_{-\infty}^{\infty}dE D_{N}(E) \cdot D_{S} (E+eV_{SD}) \cdot T_{QD}(E, V_{BG},V_{SD}) \cdot  [f_N(E) - f_S(E+eV_{SD})] $,
with $D_N(E)$ the constant DOS of the normal metal, $T_{QD}(E,V_{BG},V_{SD})$ a Lorentzian transmission function, accounting for the resonant tunneling trough the QD and including a broadening due to the finite coupling to the electrodes. \linebreak $f_{S/N}$ are the Fermi distribution functions for the respective contacts. To account for the softness of the gap, the DOS in the LS can be expressed using the phenomenological Dynes parameter $\delta$ by $ D_S = | \mathfrak{Re} (E-i\delta/\sqrt{(E-i\delta)^2-\Delta\textsuperscript{*}^2}) | $\cite{Dynes1978}.
By adjusting the magnitude of the gap $\Delta^*$, the QD resonance broadening $\Gamma$ and the Dynes parameter $\delta$, we get the conductance maps represented in the central panels in Fig.\ref{fig:Coulomb}a-d (``model'').\\
We can reproduce the characteristics of the CB resonance in  Fig. \ref{fig:Coulomb}a using $\Gamma$ = \SI{150}{\micro eV}, which is slightly smaller than what we obtained in the normal state. For the size of the gap we find $\Delta\textsuperscript{*}$ \SI{\approx165}{\micro eV}, and $\delta = 0.4\cdot \Delta\textsuperscript{*} $ ($\delta = $\SI{65}{\micro eV}), resulting in a suppression of $S\approx 0.5$, similar to what is found in the cotunneling regime. 
To reproduce the data in Fig. \ref{fig:Coulomb}b we use $\Gamma$ = \SI{40}{\micro eV} (similar to the numbers in the normal state), $\Delta\textsuperscript{*}\approx$ \SI{145}{\micro eV} and $\delta = 0.4\cdot \Delta\textsuperscript{*} $($\delta \approx$  \SI{60}{\micro eV}), i.e. the parameters are almost identical to the ones we obtained in the cotunneling regime at higher gate voltages. The corresponding cross sections agree well with the experiment. We note, that the conductance enhancement at the gap edge, as well as the negative differential conductance are reproduced by the model. We point out that in the sequential tunneling regime, we can extract gap characteristics down to back-gate voltages of 0.2V, which is not possible in the cotunneling regime.\\
%
To reproduce the characteristics of the CB resonance in Fig. \ref{fig:Coulomb}c ($V_{BG}\approx$ \SI{200}{\milli V}), we find that the tunnel coupling is similar to the one in Fig. \ref{fig:Coulomb}b $(\Gamma$ = \SI{60}{\micro eV}), and $\delta = 0.5\cdot \Delta\textsuperscript{*} $($\delta \approx$  \SI{40}{\micro eV}). However, the size of the superconducting energy gap is found to be $\Delta\textsuperscript{*}\approx$ \SI{85}{\micro eV}, which is significantly smaller than that found at larger gate voltages. Also here, the model reproduces the data well, illustrated in the corresponding cross sections.\\ 
Using the resonant tunneling model to simulate the resonance around $V_{BG}\approx$ \SI{80}{\milli V} (Fig. \ref{fig:Coulomb}d) we extract $\Gamma$ = \SI{50}{\micro eV} and an upper limit for $\Delta\textsuperscript{*}$ of \SI{10}{\micro eV}.
The model reproduces very well the characteristics of the CB resonance, which essentially corresponds to an N - QD - N device. The resonances in the experiment at higher bias outside the CB are due to excited states, which are not included in the model.\\
%
To summarize the measurements in the CB resonance regime (Fig.\ref{fig:Coulomb}): we observe a transition from a region where the LS acts as a superconducting lead (large gate voltages) with $\Delta\textsuperscript{*}\approx$ \SIrange{165}{150}{\micro eV} to an  intermediate regime with a reduced $\Delta\textsuperscript{*}\approx$ \SI{85}{\micro eV}, to a regime without effects of superconductivity.
The parameters extracted from the resonant tunneling model demonstrate a clear evolution of $\Delta\textsuperscript{*}$ in the LS.
%
\subsection{\label{sec:level5}Discussion}
\begin{figure}[!t]
	\begin{center}
	\includegraphics[width=\textwidth,height=\textheight,keepaspectratio]{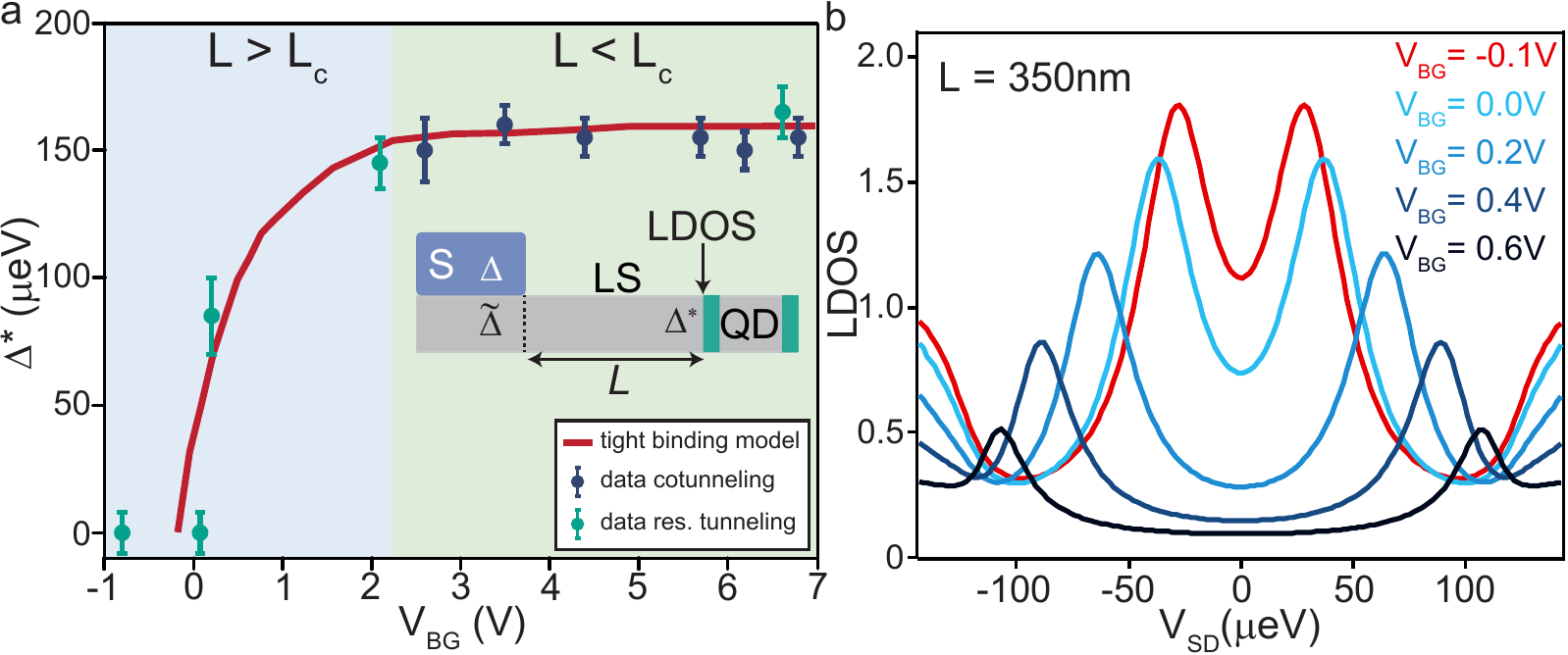}
    \end{center} 	
	\caption{\label{fig:DeltaGate} \textbf{a} Proximity gap $\Delta\textsuperscript{*}$ as a function of the gate voltage $V_{BG}$. The inset shows a schematic of the device with the relevant parameters. \textbf{b} Calculated local DOS as a function of energy for various gate voltages at a fixed distance from the interface, $L =$\SI{350}{nm}.}
\end{figure}

This evolution of the induced gap $\Delta\textsuperscript{*}$ as a function of $V_{BG}$ is summarised on Fig. \ref{fig:DeltaGate}a. The curve shows a sharp transition from a clearly resolved energy gap for $V_{BG} > 0$ to a fully suppressed gap at $V_{BG} < 0$.\\
While the observed proximity feature can well be fitted with a broadened BCS DOS, this approach is not an adequate description, since there are only few states in the quasi one-dimensional NW lead. Qualitatively, one can understand the transition by considering only a few modes in the LS. All electrons at energies within the gap of the SC are Andreev reflected (AR) at the SC, giving rise to Andreev-bound states (ABSs).\\
We interpret the observed transition in $\Delta\textsuperscript{*}$ qualitatively as a gate-tunable transition of ABSs forming in the LS from the long to the short junction limit. Both limits are defined by comparing the physical length of the junction, $L$, to the characteristic length-scale $L_{c}=\hbar v_F/\widetilde{\Delta}$, which is the coherence length in the ballistic limit\cite{Bagwell1992}. Here $v_F$ is the Fermi velocity in the LS, $\hbar$ the Planck constant and $\widetilde{\Delta}$ the proximity gap induced by the aluminium contact in the NW directly below the SC (see inset of Fig. \ref{fig:DeltaGate}a).\\
In the short junction limit ($L<<L_{c}$) the energy of the ABSs ($E_{ABS}$) in the LS is dominated by the phase change due to AR at the SC/LS interface (dashed line in schematic), where $\widetilde{\Delta}$ is assumed to change abruptly. In this limit, the energies of the ABSs are ``pushed'' to $E_{ABS} \cong \widetilde{\Delta}$, resulting in the superconducting proximity gap in the LS that is similar to $\widetilde{\Delta}$, i.e. $ \Delta^* \sim \widetilde{\Delta}$\cite{Bena2012}.
In the long junction limit ($L>>L_{c}$), $E_{ABS}$ is determined by the phase acquired in the LS, which scales with $k_{f} = m\textsuperscript{*}\cdot v_F/\hbar \propto v_F$, with $m\textsuperscript{*}$ the effective electron mass. In this limit, $E_{ABS}$ can take on smaller values, thus filling the proximity gap\cite{Bagwell1992}.
We therefore can tune $E_{ABS}$ by tuning $v_F$ and the electron density in the LS using $V_{BG}$. For very positive gate voltages, $E_F$ is relatively far up in the conduction band with a correspondingly large $v_F$, resulting in a large $L_{c}$, bringing the LS to the short junction limit, $L<L_{c}$.
In contrast, when we tune $E_F$ to the bottom of the conduction band by lowering $V_{BG}$, $v_F$ and $L_{c}$ are strongly reduced, bringing the LS to the long junction limit, $L>L_{c}$, and $E_{ABS}< \widetilde{\Delta}$. As a result, the apparent gap $\Delta\textsuperscript{*}$ is reduced by the ABSs moving into the gap. We note that $\widetilde{\Delta}$ in the NW segment located below the SC is screened by the SC and is therefore not gate tunable.\\
To support this qualitative picture, we employ a numerical model, in which we combine the Green`s function method with a tight binding model. The properties of the superconductor are taken into account  as a self energy dressing the bare Green`s function in a NW section below $S$, where a superconducting gap $\widetilde{\Delta}$  is induced depending only on the coupling to $S$ (details can be found in the supplementary materials). This region is coupled to a bare NW segment (of length $L =$ \SI{350}{\nano \meter}) modelling the LS  (see inset of Fig. \ref{fig:DeltaGate}a). We then investigate the local density of states (LDOS) at a distance $L=$\SI{350}{\nano \meter} as a function of the gate voltage in the NW (Fig. \ref{fig:DeltaGate}b). Like for the experimental data, we extract $\Delta\textsuperscript{*}$ as the distance between the maxima in the DOS. Here the number of ABSs as well as the ABS energy is determined by the total length of the LS. The resulting $\Delta\textsuperscript{*}$ of this calculation is plotted as a function of $V_{BG}$ as a red line in Fig. \ref{fig:DeltaGate}a. In excellent agreement with the experiment, we observe a sharp transition in the detected gap, corresponding to the transition from the short to the long junction limit. In the model, $\Delta\textsuperscript{*}$ tends to zero when $E_F$ is aligned with the bottom of the conduction band ($V_{BG}$ \SI{\approx -0.2}{V}), just before the LS is fully depleted and $v_F \rightarrow 0$. For a better understanding we study the dependence of an individual ABS as a function of $V_{BG}$ in the supplementary (see Fig. \ref{fig:width3Dplot2}). Evolving from the long to the short junction limit, the states with energy $E < \Delta\textsuperscript{*}$ move towards the gap, resulting in a fully formed smooth proximitised gap $\Delta\textsuperscript{*}$. We note that in this model the width of the ABS is a free parameter, which we set to \SI{25}{\micro eV}. In seeming contradiction to the depleted LS, the experiment still shows several CB resonances below this gate voltage, which we attribute to  evanescent modes from the NW segment below S (which is not depleted) that couples weakly to the QD wavefunctions, yielding the QD couplings highly asymmetric with very low transmission amplitudes, as observed in the experiment.\\
The gap in our devices is “soft”, i.e. the conductance suppression at zero bias is significantly lower than in NWs with an epitaxial Al shell\cite{Krogstrup2015,Chang2015}. The reason for this fact remains unclear  and is a priori not expected from the model. The evaporated bulk Al shows a “hard gap” ($S \approx 0.01$) when used in standard, large-area metallic S - I - S tunnel junctions measured in a similar experimental setup (see Fig.\ref{fig:SIS}). Introducing random spatial potential fluctuations at the NW-S interface\cite{Takei2013,Liu2017a} in the presented numerical model does not account for the observed small suppression, either (see Fig.\ref{fig:Disorder}). In the ABS picture, the softness of the gap is determined by the life time of the Andreev bound states in the LS. This broadening can have different physical origins, namely 1) tunneling to the QD, which should exhibit a similar tunability as the QD life time, 2) single particle tunneling to the NW segment below S, for example by an inverse proximity effect due to the gold nanoparticle used for the NW growth, and 3) quasiparticle excitation by microwave radiation absorption that might be different in NW devices compared to metallic ones.\\
In conclusion, we present a systematic study of the apparent transport gap in a NW segment induced by a proximity coupled superconductor. For this purpose we introduce QD tunnel spectroscopy enabled by in-situ grown axial tunnel barriers. We observe a gate-tunable transition of the gap amplitude from a fully developed, constant proximity gap at large electron densities to smaller values and ultimately a complete suppression of the gap at low densities. The data are consistent with a transition from the short junction limit of an S-N device to the long junction limit with ABSs forming at energies also below the gap energy. This transition occurs when the Fermi energy is close to the bottom of the conduction band and the respective Fermi velocity tends towards zero.
Our experiments demonstrate that NWs with in-situ grown barriers, in our example with crystal phase engineered barriers, are very useful to perform unambiguous transport spectroscopy in superconductor-semiconductor hybrid systems.
We have thus introduced a novel spectroscopy tool, which is well suited to study superconducting bound states in semiconducting NWs, and open, for example, an unambiguous path for battling fundamental limitations found in recent studies of Majorana bound states\cite{Liu2017,Reeg2018}.
\newpage
\begin{methods}
\section*{Fabrication}
The InAs nanowires were grown by metal-organic-vapor-phase epitaxy (MOVPE) and have an average diameter of \SI{70}{\nano\meter}. The two segments of wurtzite crystal phase forming the tunnel barriers have a thickness of \SI{30}{\nano\meter}. The zinc-blend segment in between, which defines the QD, has a width of \SI{25}{\nano\meter}.\\
The nanowires were transferred mechanically from the growth substrate to a degenerately p-doped silicon substrate with a $\text{SiO}_2$ capping layer (\SI{400}{\nano\meter}). The substrate is used as a global back gate. For the electron beam lithography we employ pre-defined markers and contact pads, made of Ti/Au (\SI{5}{\nano\meter}/\SI{45}{\nano\meter}). The normal metal contact to the NW is made of Ti/Au (\SI{5}{\nano\meter}/\SI{70}{\nano\meter}) and the superconducting contact of Ti/Al (\SI{5}{\nano\meter}/\SI{80}{\nano\meter}). Before each evaporation step, the native oxide of the NW is removed by an Argon ion sputtering.\\
All measurements were carried out in a dilution refrigerator at a base temperature of \SI{20}{\milli\kelvin}. Differential conductance has been measured using standard lock-in techniques ($V_{\text{ac}} =$ \SI{10}{\micro V}, $f_{\text{ac}} =$ \SI{278}{\hertz}).
\end{methods}
\begin{addendum}
 \item 
 
This work has received funding from the Swiss National Science Foundation, the Swiss Nanoscience Institute and the Swiss NCCR QSIT. C.S. has received funding from the European Research Council (ERC) under the European Union’s Horizon 2020 research and innovation program: grant agreement 787414, ERC-Adv TopSupra. D.C. received funding from the European Union’s Horizon 2020 research and innovation program (ERC Starting Grant, agreement No 757725). S.L., M.N., K.A.D. and C. T. acknowledge financial support by the Knut and Alice Wallenberg Foundation (KAW) and the Swedish Research Council (VR). The authors thank C. Reeg and S. Hoffman for fruitful discussions.
 
 \item[Competing Interests] The authors declare that they have no
competing financial interests.

\item[Author contributions]
C.J. fabricated the devices and performed the measurements together with R.D.. C.J., R.D. and A.B. analysed the data. A.B. provided the resonant tunneling model. D.C. provided the numerical calculations. S.L., M.N., K.A.D. and C.T. developed the nanowire structure. C.S. and A.B. planed and designed the experiments and participated in all discussions. All authors contributed to the manuscript.

 \item[Correspondence] Correspondence and requests for materials
should be addressed to C.J.~(email: christian.juenger@unibas.ch) or A.B. ~(email: andreas.baumgartner@unibas.ch).

\end{addendum}
\renewcommand{\thefigure}{S\arabic{figure}}
\setcounter{figure}{0}
\section*{Supplementary Information}
\section*{Numerical Model of a proximitised NW}
We model a 1D NW with strong spin-orbit coupling partially covered by a superconductor. The total Hamiltonian for such a system can be written as $ H= H_{NW} + H_{S} + H_{T}$. In the Nambu basis the creation operator of the NW electrons is given by $\hat{c}^\dagger = ( c_{\uparrow}^\dagger, c_{\downarrow}^\dagger, c_{\downarrow}, -c_{\uparrow} )$, where the semiconducting nanowire Hamiltonian is
\begin{equation}
H_{NW} = \int \hat{c}^\dagger\left[ (\frac{p^2}{2m}-\mu)\tau_z+\alpha p \sigma_y \tau_z\right]\hat{c}\:dx,
\end{equation}
with $\sigma$($\tau$) the Pauli matrices in the spin (particle-hole), $\mu$ the electrochemical potential and $\alpha$ the strength of the Rashba spin-orbit coupling. To be able to study the gate dependence/length dependence in our system, the Hamiltonian of the NW is described by a lattice tight-binding model where the continuum limit Hamiltonian becomes
\begin{equation}
H_{NW}=\sum^N_j c^\dagger_j \left[(t-\mu)\tau_z\right]c_j-\frac{1}{2}c^\dagger_j\left[t\tau_z+i\alpha\sigma_y\tau_z + h.c.\right]c_{j+1} ,
\end{equation}
with $c^\dagger_j$  is the creation operator of an electron in the nanowire on site $j$ written in the Nambu basis.\\
The Hamiltonian of the bulk superconductor can be written as
\begin{equation}
	H_{S}=\sum_{\mathbf k,\sigma}\xi_{k} \Psi^{\dagger}_{\mathbf k,\sigma}\Psi_{\mathbf k,\sigma}+\Delta \Psi^{\dagger}_{\mathbf k,\uparrow}\Psi^{\dagger}_{-\mathbf k,\downarrow}+h.c.	
\end{equation}
with $\xi_{k}=\frac{k^{2}}{2m}-\mu$ and $\Psi_{\mathbf k,\sigma}$ the annihilation operator for an electron in the superconductor with spin $\sigma$ and momentum $\mathbf k$. 
The tunneling Hamiltonian describing the coupling between the bulk SC and the nanowire takes on the form
\begin{equation}
	H_{T}=\sum_{j,\sigma}  \tilde{t}_j c^{\dagger}_{j,\sigma} \sum_{\mathbf k} \Psi^{\dagger}_{\mathbf k,\sigma} e^{i k_x j} + h.c.,
\end{equation}
where the operator $c^{\dagger}_{j,\sigma}$ creates an electron of spin $\sigma$ on site $j$. Since the total Hamiltonian is quadratic in the SC degrees of freedom, we can integrate them out, so that all the properties of the bulk superconductor and the tunneling to the nanowire are taken into account in the superconducting self-energy.  The total Green`s function of the system can be obtained by dressing the Green`s function of the nanowire by the superconducting self-energy such that
\begin{equation}
\tilde{G}_R^{-1} (\omega) = \tilde{G}^{-1}_{0 R} (\omega)-\tilde{\Sigma}_R^{S}(\omega).
\end{equation}
The total retarded superconducting self-energy has the following form $\tilde{\Sigma}_R^{S}(\omega)={\cal I}\otimes\tilde{\Sigma}_{j,R}^{S}(\omega)$, where ${\cal I}$ is the unity matrix in the space of sites, and $\tilde{\Sigma}_{j,R}^{S}$ the on-site retarded self-energy
\begin{equation}
\tilde{\Sigma}_{j,R}^{S} (\omega) =\left|\tilde{t}_j\right|^2 \tau_z~\tilde{g}_R(\omega)~ \tau_z,
\end{equation}
$\tilde{g}_R(\omega)$ the Green`s function of the superconductor. This Green`s function is well known \cite{Yeyati96,Chevallier2011} and can be incorporated using the superconducting self-energy
\begin{equation}\hat{\Sigma}_{j,R}^S(\omega)=\Gamma_{j,S}\tilde{g}_R(\omega)=\Gamma_{j,S} \frac{\omega}{\sqrt{\Delta^2-\omega^2}}\left({\bf 1}+\frac{\Delta}{\omega} \tau_x\right)
\end{equation}
where {\bf 1} is the unity matrix in spin and particle-hole space, and $\Gamma_{j,S}=\pi\nu(0)\left|\tilde{t}_j\right|^2$ are the tunneling rates. The retarded bare Green's function of the semiconducting NW electrons is obtained by inverting the NW Hamiltonian
\begin{equation}
\tilde{G}_{0 R}^{-1}(\omega)=(\omega+i \delta){\bf 1}-H_{NW}
\end{equation}
where $\delta$ is an infinitesimal quasiparticle inverse lifetime introduced to avoid diverging terms in the numerical evaluations. This quantity determines the width of the bound states in the system.\\
We are interested in getting the local density of states evaluated at a given site $j$ in the NW. Such a quantity can be obtained by taking the imaginary part of the total Green`s function
\begin{equation}
LDOS=-\frac{1}{\pi}\sum_{\beta=\uparrow,\downarrow}\textrm{Im}[\tilde{G}^{jj,\beta\beta}_{R}(\omega)].
\end{equation}
The hopping amplitude is defined as $t=\hbar^2/2m^*a^2$ with $m^* = 0.0025m_e$ being the effective mass, $a$ the lattice constant, and $\hbar$ the Planck constant. The relation between the experimental and the tight binding parameters are determined by $t$. We use the maximum value of the gap $\Delta^*=$\SI{160}{\micro eV}, in the short junction limit when it is fully formed, which corresponds to $t\approx1.6meV$ and a lattice constant of \SI{70}{nm}. With these values, we can fully determine the parameters in the experiment. The only free parameters in our model is the lever arm which is taken to be \SI{0.1}{\milli eV /\milli V} similar as in the experiment (\SI{0.06}{\milli eV /\milli V}). For the spin-orbit interaction strength we used \SI{70}{\micro eV}, similar to what has been reported before.
\section*{Extracted QD and LS characteristics}
\begin{figure}[!h]
	\begin{center}
	\includegraphics[width=0.99\columnwidth]{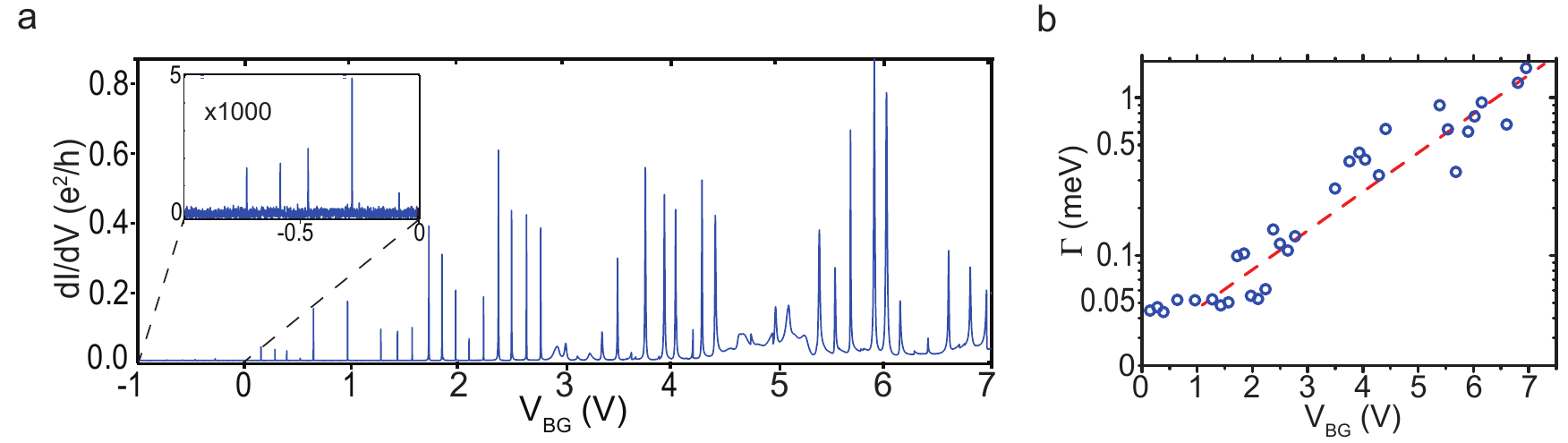}
    \end{center} 	
	\caption{Gate voltage dependence of characteristic parameters. \textbf{a} Differential conductance as a function of $V_{BG}$. \textbf{b} Logarithmic plot of QD resonance width $\Gamma$ (from Lorentzian fits) as a function of $V_{BG}$. }
	\label{fig:GammaGate}
\end{figure}

\begin{figure}[!h]
	\begin{center}
	\includegraphics[width=0.99\columnwidth]{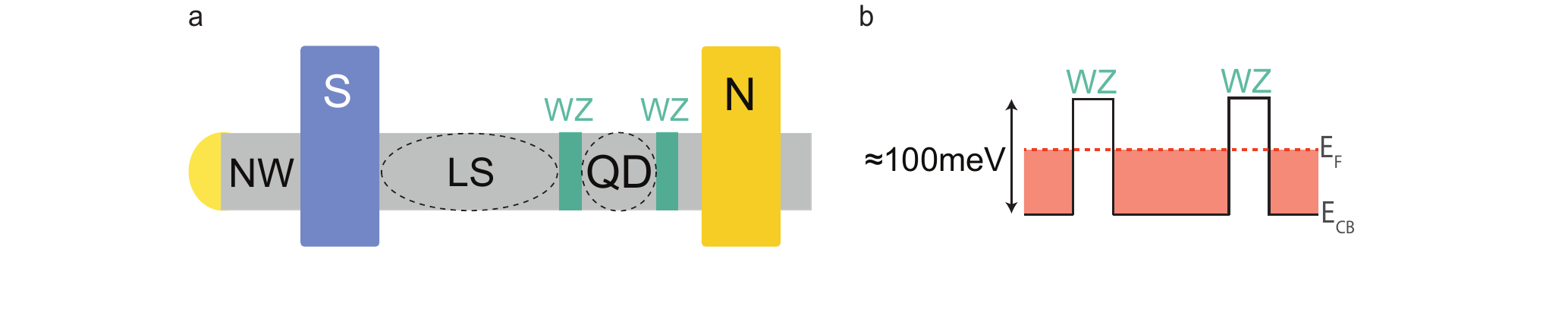}
    \end{center} 	
	\caption{Schematic of the device structure and the band structure. \textbf{a} Schematic of the NW S- N device structure \textbf{b} Illustration of the energy diagram of the conduction band $E_{CB}$ relative to the Fermi energy $E_{F}$.}
	\label{fig:EnergySchemeFermi}
\end{figure}

Fig.\ref{fig:GammaGate}a shows regular CB resonances over a large backgate voltage range in the normal state ($B =$ \SI{50}{\milli \tesla}) at $V_{SD}$=\SI{0}{}. A systematic increase of the resonance width is observed, with a strong variation between neighbouring charge states, possibly due to orbital effects. The average broadening $\Gamma$ increases with increasing $V_{BG}$ (see Fig.\ref{fig:GammaGate} b), consistent with a exponential increase with a linear reduction of the barrier height with respect to $E_F$.
\\
In Fig.\ref{fig:EnergySchemeFermi}a the schematic of the device structure is shown. The WZ tunnel barriers, defining the QD are represented in green.  Fig.\ref{fig:EnergySchemeFermi}b shows the energy diagram of the conduction band edge $E_{CB}$ relative to the Fermi energy $E_{F}$ (dashed line) with an offset of $\approx$\SI{100}{\milli eV} between WZ and ZB crystal phase.

\begin{figure}[!h]
	\begin{center}
	\includegraphics[width=0.99\columnwidth]{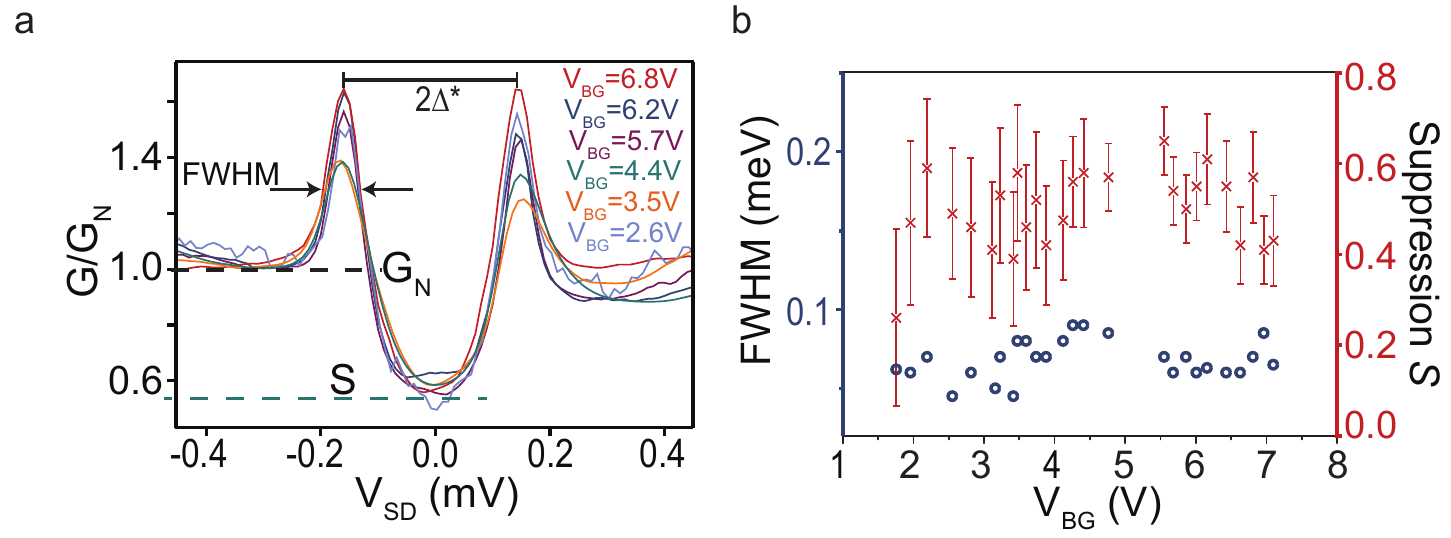}
    \end{center} 	
	\caption{\textbf{a} Differential conductance $G$ normalised to $G_N$ as a function of $V_{SD}$ at constant backgate voltages in the cotunneling regime. \textbf{b} Full Width Half maximum (blue) of the coherence peaks and normalized zero bias suppression $S$ (red) as a function $V_{BG}$. }
	\label{fig:NormGap}
\end{figure}
\clearpage
\section*{Additional numerical calculations for an individual ABS}
\begin{figure}[!b]
	\begin{center}
	\includegraphics[width=0.95\columnwidth]{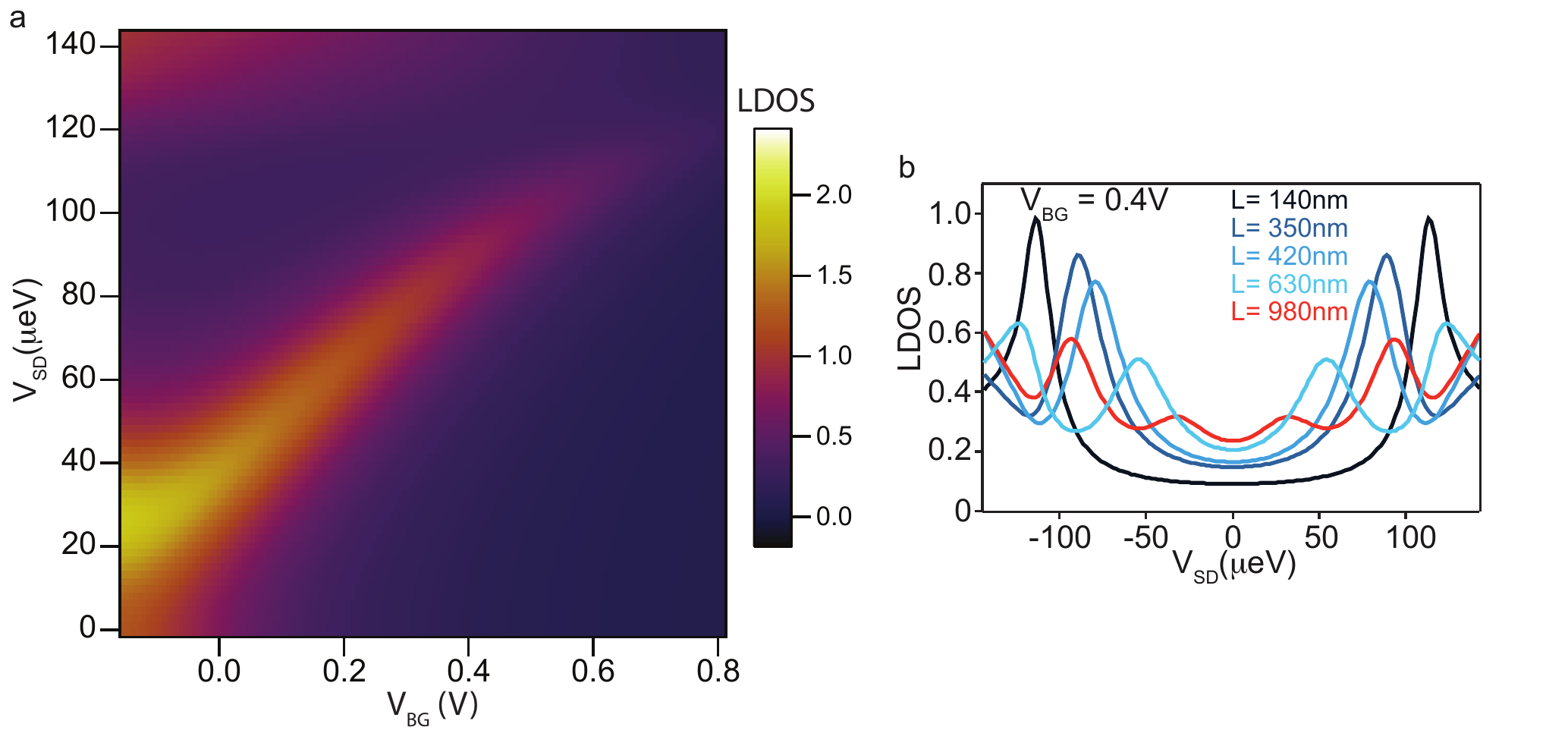}
    \end{center} 		\caption{ \textbf{a} Local density of states as a function of bias $V_{SD}$ and backgate voltage $V_{BG}$ at $L =$ \SI{350}{\nano \meter}. The ABSs width is set to $\Gamma_{\text{ABS}}=$ \SI{25}{\micro eV}. \textbf{b} LDOS as a function of energy at various distances from the interface for a fixed gate voltage $V_{BG}=0.4V$.}	
    \label{fig:width3Dplot2}
\end{figure}

Fig.\ref{fig:width3Dplot2}a shows the LDOS of an individual ABS as a function of the backgate voltage $V_{BG}$ and source drain bias $V_{SD}$ for a fixed junction length of $L =$ \SI{350}{\nano \meter}. Fig. \ref{fig:width3Dplot2}b shows cross sections for the LDOS of an individual ABS for a fixed backgate voltage $V_{BG}=0.4V$ and different lengths $L$.
\section*{Additional measurements of S - I - S junction}
\begin{figure}[!h]
	\begin{center}
	\includegraphics[width=0.95\columnwidth]{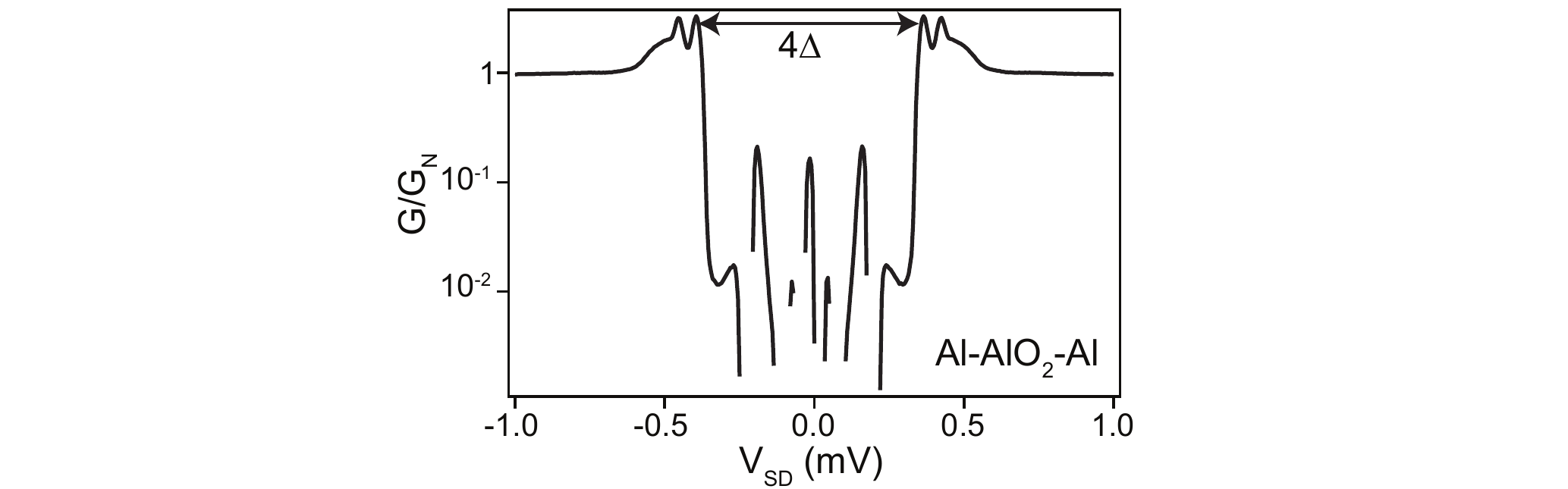}
    \end{center} 		\caption{  \textbf{a} Normalised conductance $G/G_N$ as a function of $V_{SD}$ of an S-I-S tunnel junction (made of Al- Al$\text{O}_2$-Al).}	
    \label{fig:SIS}
\end{figure}

Fig.\ref{fig:SIS} shows the normalized $G$ of a S-I-S tunnel junction (consisting of Al-Al$\text{O}_2$-Al). The sample was measured in a similar measurement setup as the one discussed in the main text.\\
\section*{Additional numerical calculations for interface with disorder}
\begin{figure}[!t]
	\begin{center}
	\includegraphics[width=0.95\columnwidth]{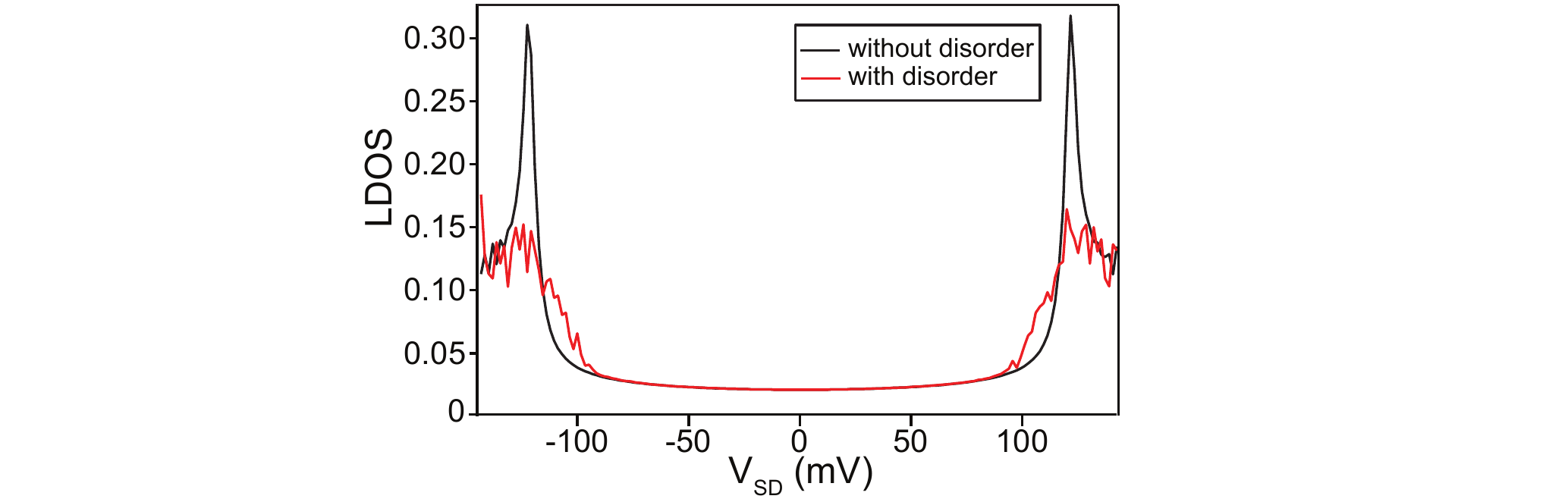}
    \end{center} 		\caption{Local density of states as a function of bias $V_{SD}$ without (black) and with (red) disorder. The disorder LDOS was averaged 30 times.}	
    \label{fig:Disorder}
\end{figure}

Fig.\ref{fig:Disorder} shows the LDOS as a function of the source drain bias $V_{SD}$  (at $V_{BG}=$\SI{1}{V}) at a distance of $L =$ \SI{70}{\nano \meter} from the S - N interface. The width of the ABS is set to $\Gamma_{\text{ABS}}=$ \SI{4}{\micro eV}. The black cross section shows the LDOS without disorder at the interface. The red cross section shows the same cross section as before, but with a random potential fluctuation at the interface. This was realized by adding a random fluctuation (maximum +/- \SI{130}{\micro eV}) to the coupling of the superconductor ($\Gamma_{S,i} =$ \SI{320}{\micro eV}) at each site $i$ of the NW below. The randomized fluctuation value follows a Gaussian distribution. The plotted cross section was averaged 30 times.  As a result, the peaks at the gap edge are smaller in amplitude and also broadened, compared to the clean case. The edges of the gap appear to be slightly smoother.
\clearpage
\newpage
\newpage
\section*{References}
\bibliography{literature}

\begin{thebibliography}{10}
\expandafter\ifx\csname url\endcsname\relax
  \def\url#1{\texttt{#1}}\fi
\expandafter\ifx\csname urlprefix\endcsname\relax\def\urlprefix{URL }\fi
\providecommand{\bibinfo}[2]{#2}
\providecommand{\eprint}[2][]{\url{#2}}

\bibitem{Beenakker1992}
\bibinfo{author}{Beenakker, C. W.~J.}
\newblock \bibinfo{title}{Quantum transport in semiconductor-superconductor
  microjunctions}.
\newblock \emph{\bibinfo{journal}{Physical Review B}}
  \textbf{\bibinfo{volume}{46}}, \bibinfo{pages}{12841--12844}
  (\bibinfo{year}{1992}).

\bibitem{Hofstetter2009}
\bibinfo{author}{Hofstetter, L.}, \bibinfo{author}{Csonka, S.},
  \bibinfo{author}{Nyg{\aa}rd, J.} \& \bibinfo{author}{Sch\"onenberger, C.}
\newblock \bibinfo{title}{Cooper pair splitter realized in a two-quantum-dot
  y-junction}.
\newblock \emph{\bibinfo{journal}{Nature}} \textbf{\bibinfo{volume}{461}},
  \bibinfo{pages}{960--963} (\bibinfo{year}{2009}).

\bibitem{Schindele2012}
\bibinfo{author}{Schindele, J.}, \bibinfo{author}{Baumgartner, A.} \&
  \bibinfo{author}{Schönenberger, C.}
\newblock \bibinfo{title}{Near-unity cooper pair splitting efficiency}.
\newblock \emph{\bibinfo{journal}{Physical Review Letters}}
  \textbf{\bibinfo{volume}{109}} (\bibinfo{year}{2012}).

\bibitem{Cleuziou2006}
\bibinfo{author}{Cleuziou, J.-P.}, \bibinfo{author}{Wernsdorfer, W.},
  \bibinfo{author}{Bouchiat, V.}, \bibinfo{author}{Ondar{\c{c}}uhu, T.} \&
  \bibinfo{author}{Monthioux, M.}
\newblock \bibinfo{title}{Carbon nanotube superconducting quantum interference
  device}.
\newblock \emph{\bibinfo{journal}{Nature Nanotechnology}}
  \textbf{\bibinfo{volume}{1}}, \bibinfo{pages}{53--59} (\bibinfo{year}{2006}).

\bibitem{Mourik2012}
\bibinfo{author}{Mourik, V.} \emph{et~al.}
\newblock \bibinfo{title}{Signatures of majorana fermions in hybrid
  superconductor-semiconductor nanowire devices}.
\newblock \emph{\bibinfo{journal}{Science}} \textbf{\bibinfo{volume}{336}},
  \bibinfo{pages}{1003--1007} (\bibinfo{year}{2012}).

\bibitem{Albrecht2016}
\bibinfo{author}{Albrecht, S.~M.} \emph{et~al.}
\newblock \bibinfo{title}{Exponential protection of zero modes in majorana
  islands}.
\newblock \emph{\bibinfo{journal}{Nature}} \textbf{\bibinfo{volume}{531}},
  \bibinfo{pages}{206--209} (\bibinfo{year}{2016}).

\bibitem{Deng2016}
\bibinfo{author}{Deng, M.~T.} \emph{et~al.}
\newblock \bibinfo{title}{Majorana bound state in a coupled quantum-dot
  hybrid-nanowire system}.
\newblock \emph{\bibinfo{journal}{Science}} \textbf{\bibinfo{volume}{354}},
  \bibinfo{pages}{1557--1562} (\bibinfo{year}{2016}).

\bibitem{Zhang2018}
\bibinfo{author}{Zhang, H.} \emph{et~al.}
\newblock \bibinfo{title}{Quantized majorana conductance}.
\newblock \emph{\bibinfo{journal}{Nature}} \textbf{\bibinfo{volume}{556}},
  \bibinfo{pages}{74--79} (\bibinfo{year}{2018}).

\bibitem{Alicea2011}
\bibinfo{author}{Alicea, J.}, \bibinfo{author}{Oreg, Y.},
  \bibinfo{author}{Refael, G.}, \bibinfo{author}{von Oppen, F.} \&
  \bibinfo{author}{Fisher, M. P.~A.}
\newblock \bibinfo{title}{Non-abelian statistics and topological quantum
  information processing in 1d wire networks}.
\newblock \emph{\bibinfo{journal}{Nature Physics}}
  \textbf{\bibinfo{volume}{7}}, \bibinfo{pages}{412--417}
  (\bibinfo{year}{2011}).

\bibitem{Nayak2008}
\bibinfo{author}{Nayak, C.}, \bibinfo{author}{Simon, S.~H.},
  \bibinfo{author}{Stern, A.}, \bibinfo{author}{Freedman, M.} \&
  \bibinfo{author}{Sarma, S.~D.}
\newblock \bibinfo{title}{Non-abelian anyons and topological quantum
  computation}.
\newblock \emph{\bibinfo{journal}{Reviews of Modern Physics}}
  \textbf{\bibinfo{volume}{80}}, \bibinfo{pages}{1083--1159}
  (\bibinfo{year}{2008}).

\bibitem{Leijnse2011}
\bibinfo{author}{Leijnse, M.} \& \bibinfo{author}{Flensberg, K.}
\newblock \bibinfo{title}{Scheme to measure majorana fermion lifetimes using a
  quantum dot}.
\newblock \emph{\bibinfo{journal}{Physical Review B}}
  \textbf{\bibinfo{volume}{84}} (\bibinfo{year}{2011}).

\bibitem{Hoffman2017}
\bibinfo{author}{Hoffman, S.}, \bibinfo{author}{Chevallier, D.},
  \bibinfo{author}{Loss, D.} \& \bibinfo{author}{Klinovaja, J.}
\newblock \bibinfo{title}{Spin-dependent coupling between quantum dots and
  topological quantum wires}.
\newblock \emph{\bibinfo{journal}{Physical Review B}}
  \textbf{\bibinfo{volume}{96}} (\bibinfo{year}{2017}).

\bibitem{Chevallier2018}
\bibinfo{author}{Chevallier, D.}, \bibinfo{author}{Szumniak, P.},
  \bibinfo{author}{Hoffman, S.}, \bibinfo{author}{Loss, D.} \&
  \bibinfo{author}{Klinovaja, J.}
\newblock \bibinfo{title}{Topological phase detection in rashba nanowires with
  a quantum dot}.
\newblock \emph{\bibinfo{journal}{Physical Review B}}
  \textbf{\bibinfo{volume}{97}} (\bibinfo{year}{2018}).

\bibitem{Gharavi2016}
\bibinfo{author}{Gharavi, K.}, \bibinfo{author}{Hoving, D.} \&
  \bibinfo{author}{Baugh, J.}
\newblock \bibinfo{title}{Readout of majorana parity states using a quantum
  dot}.
\newblock \emph{\bibinfo{journal}{Physical Review B}}
  \textbf{\bibinfo{volume}{94}} (\bibinfo{year}{2016}).

\bibitem{Dick2010}
\bibinfo{author}{Dick, K.~A.}, \bibinfo{author}{Thelander, C.},
  \bibinfo{author}{Samuelson, L.} \& \bibinfo{author}{Caroff, P.}
\newblock \bibinfo{title}{Crystal phase engineering in single {InAs}
  nanowires}.
\newblock \emph{\bibinfo{journal}{Nano Letters}} \textbf{\bibinfo{volume}{10}},
  \bibinfo{pages}{3494--3499} (\bibinfo{year}{2010}).

\bibitem{Lehmann2013}
\bibinfo{author}{Lehmann, S.}, \bibinfo{author}{Wallentin, J.},
  \bibinfo{author}{Jacobsson, D.}, \bibinfo{author}{Deppert, K.} \&
  \bibinfo{author}{Dick, K.~A.}
\newblock \bibinfo{title}{A general approach for sharp crystal phase switching
  in {InAs}, {GaAs}, {InP}, and {GaP} nanowires using only group v flow}.
\newblock \emph{\bibinfo{journal}{Nano Letters}} \textbf{\bibinfo{volume}{13}},
  \bibinfo{pages}{4099--4105} (\bibinfo{year}{2013}).

\bibitem{Rossella2014}
\bibinfo{author}{Rossella, F.} \emph{et~al.}
\newblock \bibinfo{title}{Nanoscale spin rectifiers controlled by the stark
  effect}.
\newblock \emph{\bibinfo{journal}{Nature Nanotechnology}}
  \textbf{\bibinfo{volume}{9}}, \bibinfo{pages}{997--1001}
  (\bibinfo{year}{2014}).

\bibitem{Nilsson2018}
\bibinfo{author}{Nilsson, M.} \emph{et~al.}
\newblock \bibinfo{title}{Tuning the two-electron hybridization and spin states
  in parallel-coupled {InAs} quantum dots}.
\newblock \emph{\bibinfo{journal}{Physical Review Letters}}
  \textbf{\bibinfo{volume}{121}} (\bibinfo{year}{2018}).

\bibitem{Nilsson2016}
\bibinfo{author}{Nilsson, M.} \emph{et~al.}
\newblock \bibinfo{title}{Single-electron transport in {InAs} nanowire quantum
  dots formed by crystal phase engineering}.
\newblock \emph{\bibinfo{journal}{Physical Review B}}
  \textbf{\bibinfo{volume}{93}} (\bibinfo{year}{2016}).

\bibitem{Chen2017}
\bibinfo{author}{Chen, I.-J.} \emph{et~al.}
\newblock \bibinfo{title}{Conduction band offset and polarization effects in
  {InAs} nanowire polytype junctions}.
\newblock \emph{\bibinfo{journal}{Nano Letters}} \textbf{\bibinfo{volume}{17}},
  \bibinfo{pages}{902--908} (\bibinfo{year}{2017}).

\bibitem{Doh2008}
\bibinfo{author}{Doh, Y.-J.}, \bibinfo{author}{Franceschi, S.~D.},
  \bibinfo{author}{Bakkers, E. P. A.~M.} \& \bibinfo{author}{Kouwenhoven,
  L.~P.}
\newblock \bibinfo{title}{Andreev reflection versus coulomb blockade in hybrid
  semiconductor nanowire devices}.
\newblock \emph{\bibinfo{journal}{Nano Letters}} \textbf{\bibinfo{volume}{8}},
  \bibinfo{pages}{4098--4102} (\bibinfo{year}{2008}).

\bibitem{Gramich2015}
\bibinfo{author}{Gramich, J.}, \bibinfo{author}{Baumgartner, A.} \&
  \bibinfo{author}{Sch\"onenberger, C.}
\newblock \bibinfo{title}{Resonant and inelastic andreev tunneling observed on
  a carbon nanotube quantum dot}.
\newblock \emph{\bibinfo{journal}{Physical Review Letters}}
  \textbf{\bibinfo{volume}{115}} (\bibinfo{year}{2015}).

\bibitem{Gramich2017}
\bibinfo{author}{Gramich, J.}, \bibinfo{author}{Baumgartner, A.} \&
  \bibinfo{author}{Sch\"onenberger, C.}
\newblock \bibinfo{title}{Andreev bound states probed in three-terminal quantum
  dots}.
\newblock \emph{\bibinfo{journal}{Physical Review B}}
  \textbf{\bibinfo{volume}{96}} (\bibinfo{year}{2017}).

\bibitem{Court2007}
\bibinfo{author}{Court, N.~A.}, \bibinfo{author}{Ferguson, A.~J.} \&
  \bibinfo{author}{Clark, R.~G.}
\newblock \bibinfo{title}{Energy gap measurement of nanostructured aluminium
  thin films for single cooper-pair devices}.
\newblock \emph{\bibinfo{journal}{Superconductor Science and Technology}}
  \textbf{\bibinfo{volume}{21}}, \bibinfo{pages}{015013}
  (\bibinfo{year}{2007}).

\bibitem{Franceschi2001}
\bibinfo{author}{Franceschi, S.~D.} \emph{et~al.}
\newblock \bibinfo{title}{Electron cotunneling in a semiconductor quantum dot}.
\newblock \emph{\bibinfo{journal}{Physical Review Letters}}
  \textbf{\bibinfo{volume}{86}}, \bibinfo{pages}{878--881}
  (\bibinfo{year}{2001}).

\bibitem{Doh2005}
\bibinfo{author}{Doh, Y.-J.}
\newblock \bibinfo{title}{Tunable supercurrent through semiconductor
  nanowires}.
\newblock \emph{\bibinfo{journal}{Science}} \textbf{\bibinfo{volume}{309}},
  \bibinfo{pages}{272--275} (\bibinfo{year}{2005}).

\bibitem{Sand-Jespersen2007}
\bibinfo{author}{Sand-Jespersen, T.} \emph{et~al.}
\newblock \bibinfo{title}{Kondo-enhanced andreev tunneling in {InAs} nanowire
  quantum dots}.
\newblock \emph{\bibinfo{journal}{Physical Review Letters}}
  \textbf{\bibinfo{volume}{99}} (\bibinfo{year}{2007}).

\bibitem{Jellinggaard2016}
\bibinfo{author}{Jellinggaard, A.}, \bibinfo{author}{Grove-Rasmussen, K.},
  \bibinfo{author}{Madsen, M.~H.} \& \bibinfo{author}{Nyg{\aa}rd, J.}
\newblock \bibinfo{title}{Tuning yu-shiba-rusinov states in a quantum dot}.
\newblock \emph{\bibinfo{journal}{Physical Review B}}
  \textbf{\bibinfo{volume}{94}} (\bibinfo{year}{2016}).

\bibitem{Gramich2016}
\bibinfo{author}{Gramich, J.}, \bibinfo{author}{Baumgartner, A.} \&
  \bibinfo{author}{Sch\"onenberger, C.}
\newblock \bibinfo{title}{Subgap resonant quasiparticle transport in
  normal-superconductor quantum dot devices}.
\newblock \emph{\bibinfo{journal}{Applied Physics Letters}}
  \textbf{\bibinfo{volume}{108}}, \bibinfo{pages}{172604}
  (\bibinfo{year}{2016}).

\bibitem{Yeyati1997}
\bibinfo{author}{Yeyati, A.~L.}, \bibinfo{author}{Cuevas, J.~C.},
  \bibinfo{author}{L{\'{o}}pez-D{\'{a}}valos, A.} \&
  \bibinfo{author}{Mart{\'{\i}}n-Rodero, A.}
\newblock \bibinfo{title}{Resonant tunneling through a small quantum dot
  coupled to superconducting leads}.
\newblock \emph{\bibinfo{journal}{Physical Review B}}
  \textbf{\bibinfo{volume}{55}}, \bibinfo{pages}{R6137--R6140}
  (\bibinfo{year}{1997}).

\bibitem{Dynes1978}
\bibinfo{author}{Dynes, R.~C.}, \bibinfo{author}{Narayanamurti, V.} \&
  \bibinfo{author}{Garno, J.~P.}
\newblock \bibinfo{title}{Direct measurement of quasiparticle-lifetime
  broadening in a strong-coupled superconductor}.
\newblock \emph{\bibinfo{journal}{Physical Review Letters}}
  \textbf{\bibinfo{volume}{41}}, \bibinfo{pages}{1509--1512}
  (\bibinfo{year}{1978}).

\bibitem{Bagwell1992}
\bibinfo{author}{Bagwell, P.~F.}
\newblock \bibinfo{title}{Suppression of the josephson current through a
  narrow, mesoscopic, semiconductor channel by a single impurity}.
\newblock \emph{\bibinfo{journal}{Physical Review B}}
  \textbf{\bibinfo{volume}{46}}, \bibinfo{pages}{12573--12586}
  (\bibinfo{year}{1992}).

\bibitem{Bena2012}
\bibinfo{author}{Bena, C.}
\newblock \bibinfo{title}{Metamorphosis and taxonomy of andreev bound states}.
\newblock \emph{\bibinfo{journal}{The European Physical Journal B}}
  \textbf{\bibinfo{volume}{85}} (\bibinfo{year}{2012}).

\bibitem{Krogstrup2015}
\bibinfo{author}{Krogstrup, P.} \emph{et~al.}
\newblock \bibinfo{title}{Epitaxy of semiconductor{\textendash}superconductor
  nanowires}.
\newblock \emph{\bibinfo{journal}{Nature Materials}}
  \textbf{\bibinfo{volume}{14}}, \bibinfo{pages}{400--406}
  (\bibinfo{year}{2015}).

\bibitem{Chang2015}
\bibinfo{author}{Chang, W.} \emph{et~al.}
\newblock \bibinfo{title}{Hard gap in epitaxial
  semiconductor{\textendash}superconductor nanowires}.
\newblock \emph{\bibinfo{journal}{Nature Nanotechnology}}
  \textbf{\bibinfo{volume}{10}}, \bibinfo{pages}{232--236}
  (\bibinfo{year}{2015}).

\bibitem{Takei2013}
\bibinfo{author}{Takei, S.}, \bibinfo{author}{Fregoso, B.~M.},
  \bibinfo{author}{Hui, H.-Y.}, \bibinfo{author}{Lobos, A.~M.} \&
  \bibinfo{author}{Sarma, S.~D.}
\newblock \bibinfo{title}{Soft superconducting gap in semiconductor majorana
  nanowires}.
\newblock \emph{\bibinfo{journal}{Physical Review Letters}}
  \textbf{\bibinfo{volume}{110}} (\bibinfo{year}{2013}).

\bibitem{Liu2017a}
\bibinfo{author}{Liu, C.-X.}, \bibinfo{author}{Setiawan, F.},
  \bibinfo{author}{Sau, J.~D.} \& \bibinfo{author}{Sarma, S.~D.}
\newblock \bibinfo{title}{Phenomenology of the soft gap, zero-bias peak, and
  zero-mode splitting in ideal majorana nanowires}.
\newblock \emph{\bibinfo{journal}{Physical Review B}}
  \textbf{\bibinfo{volume}{96}} (\bibinfo{year}{2017}).

\bibitem{Liu2017}
\bibinfo{author}{Liu, C.-X.}, \bibinfo{author}{Sau, J.~D.},
  \bibinfo{author}{Stanescu, T.~D.} \& \bibinfo{author}{Sarma, S.~D.}
\newblock \bibinfo{title}{Andreev bound states versus majorana bound states in
  quantum dot-nanowire-superconductor hybrid structures: Trivial versus
  topological zero-bias conductance peaks}.
\newblock \emph{\bibinfo{journal}{Physical Review B}}
  \textbf{\bibinfo{volume}{96}} (\bibinfo{year}{2017}).

\bibitem{Reeg2018}
\bibinfo{author}{Reeg, C.}, \bibinfo{author}{Dmytruk, O.},
  \bibinfo{author}{Chevallier, D.}, \bibinfo{author}{Loss, D.} \&
  \bibinfo{author}{Klinovaja, J.}
\newblock \bibinfo{title}{Zero-energy andreev bound states from quantum dots in
  proximitized rashba nanowires}.
\newblock \emph{\bibinfo{journal}{Physical Review B}}
  \textbf{\bibinfo{volume}{98}}, \bibinfo{pages}{245407}
  (\bibinfo{year}{2018}).

\bibitem{Yeyati96}
\bibinfo{author}{Cuevas, J.~C.}, \bibinfo{author}{Mart{\'{\i}}n-Rodero, A.} \&
  \bibinfo{author}{Yeyati, A.~L.}
\newblock \bibinfo{title}{Hamiltonian approach to the transport properties of
  superconducting quantum point contacts}.
\newblock \emph{\bibinfo{journal}{Physical Review B}}
  \textbf{\bibinfo{volume}{54}}, \bibinfo{pages}{7366--7379}
  (\bibinfo{year}{1996}).

\bibitem{Chevallier2011}
\bibinfo{author}{Chevallier, D.}, \bibinfo{author}{Rech, J.},
  \bibinfo{author}{Jonckheere, T.} \& \bibinfo{author}{Martin, T.}
\newblock \bibinfo{title}{Current and noise correlations in a double-dot
  cooper-pair beam splitter}.
\newblock \emph{\bibinfo{journal}{Physical Review B}}
  \textbf{\bibinfo{volume}{83}} (\bibinfo{year}{2011}).

\end{thebibliography}
\end{document}